\documentclass[12pt,epsfig]{article}
\usepackage{graphicx,amsmath,amssymb}

\newlength{\dinwidth}
\newlength{\dinmargin}
\def\lapproxeq{\lower .7ex\hbox{$\;\stackrel{\textstyle
<}{\sim}\;$}}
\def\gapproxeq{\lower .7ex\hbox{$\;\stackrel{\textstyle
>}{\sim}\;$}}
\def\gtrsim{\lower .7ex\hbox{$\;\stackrel{\textstyle
>}{\sim}\;$}}
\def\lesim{\lower .7ex\hbox{$\;\stackrel{\textstyle
<}{\sim}\;$}}

\def\be{\begin{equation}}
\def\ee{\end{equation}}
\def\bea{\begin{eqnarray}}
\def\eea{\end{eqnarray}}

\def\GeV{\rm GeV}
\def\J{J/\psi}

\begin{document}
\begin{flushright}
IPPP/06/65 \\
DCPT/06/130 \\
29th September 2006 \\

\end{flushright}

\vspace*{0.5cm}

\begin{center}
{\Large \bf The extraction of the bare triple-Pomeron vertex; a crucial ingredient for diffraction}

\vspace*{1cm}
\textsc{V.A.~Khoze$^{a,b}$, A.D. Martin$^a$ and M.G. Ryskin$^{a,b}$} \\

\vspace*{0.5cm}
$^a$ Department of Physics and Institute for
Particle Physics Phenomenology, \\
University of Durham, DH1 3LE, UK \\
$^b$ Petersburg Nuclear Physics Institute, Gatchina,
St.~Petersburg, 188300, Russia \\

\end{center}

\vspace*{0.5cm}

\begin{abstract}
The triple-Pomeron coupling lies at the heart of the predictions for high energy diffractive processes. We explain why the existing determinations, which use single-particle inclusive hadronic data $a+b \to a+Y$, underestimate the value of the bare coupling, due to the neglect of soft rescattering which populates the rapidity gaps. We describe how data for the process $\gamma+p \to \J +Y$ can be used to give a much more reliable estimate of the bare coupling. We use the existing, fragmentary, $\J$ data from HERA to show that the triple-Pomeron coupling is probably about 3 times larger than the previous determinations. We emphasize the importance of an explicit measurement of the mass spectrum of the $Y$ system which accompanies $\J$ production at HERA.
The consequences for ultra high energy cosmic ray showers are mentioned.
\end{abstract}

The triple-Pomeron coupling is a crucial ingredient for understanding diffraction. The value and the $t$-dependence of the vertex determine the asymptotic high energy behaviour of total cross sections and the cross section of diffractive dissociation. Moreover, in turn, the calculation of the rapidity gap survival factors, $S^2$, for diffractive processes depend on the properties of the triple-Pomeron vertex. The factor $S^2$ depends on the particular diffractive process, as well as the values of its kinematic variables, and is found to be of size from about 0.01 to 1. An example of an analysis of soft high energy hadronic scattering, with the consequent calculation of the $S^2$ factors, and their relation to the triple-Pomeron vertex can be found in \cite{KMRsoft}.  

\begin{figure}
\begin{center}
\includegraphics[height=5cm]{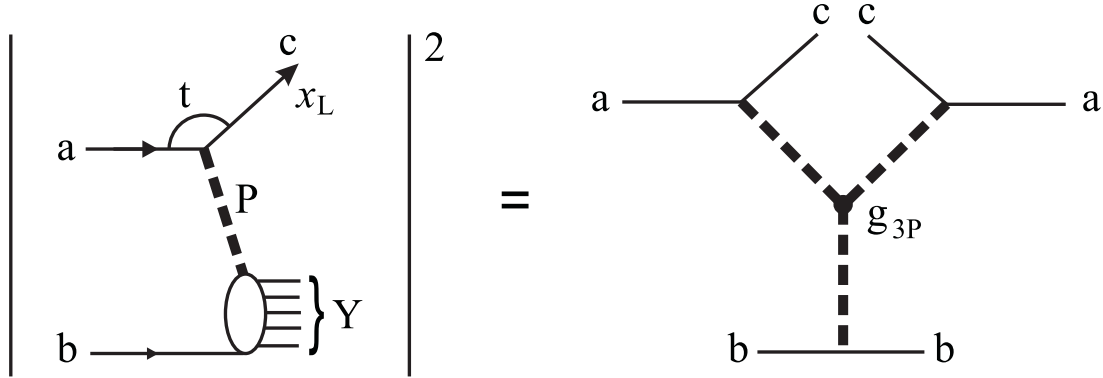}
\caption{The diffractive single-particle inclusive hadronic process $a+b \to c+Y$, which is described by a diagram with a triple-Pomeron vertex. We take $c=a$, which is always the case for the observed triple-Pomeron processes.}
\label{fig:1}
\end{center}
\end{figure}

The existing determinations of the vertex, which were made many years ago\footnote{See, for example, \cite{old}, or the review \cite{krev}}, were based on the analyses of the cross sections for single particle hadronic inclusive processes
\be
a+b ~\to~c+Y,
\label{eq:hh}
\ee
in the triple Pomeron region where the leading hadron, $c$, carries a large fraction $x_L$ of the incoming momentum, see
Fig.~\ref{fig:1}. The production of the leading hadron with $x_L \to 1$ selects Pomeron exchange with trajectory $\alpha_P(t)$, providing the quantum numbers of hadrons $a$ and $c$ are the same. For simplicity we take $c=a$, which is generally the case. In addition the production of a system $Y$ of high mass $M_Y$ is described by Pomeron exchange with trajectory $\alpha_P(0)$; leading to a diagram with a triple-Pomeron vertex, which is shown on the right-hand-side of Fig.~\ref{fig:1}. In this formalism the $a+b \to a+Y$ cross section is given by \cite{triple,collins}
\be
M_Y^2 \frac{d\sigma^{\rm SD}}{dt dM_Y^2}~=~\frac{S^2_{\rm SD}}{16\pi^2}g_{a}^2(t) g_b(0) g_{3P}(t) \left(\frac{M_Y^2}{s}\right)^{1+\alpha_P(0)-2\alpha_P(t)} \left(\frac{s}{s_0}\right)^{\alpha_P(0)-1}~~~+{\rm RRP+PPR~terms}
\label{eq:trf}
\ee
with $s_0 \equiv 1~\GeV^2$ and where $\sqrt{s} \equiv W$ is the centre-of-mass energy of the incoming $ab$ system. Here we use SD to denote single-particle diffractive dissociation. The triple-Pomeron term (PPP) is shown explicitly in (\ref{eq:trf}). There are also contributions RRP, PPR,... in which secondary Reggeons R replace the upper and/or lower Pomeron exchanges. The RRP term may give some contribution as $x_L$ decreases away from 1, and the PPR term may give some effect if the mass $M_Y$ is not sufficiently large.

Unfortunately, the determinations of the triple-Pomeron coupling $g_{3P}(0)$ from data for the hadronic $a+b \to c+Y$ processes \cite{old} are, at best, just estimates. The problem is that no account is taken of the soft rescattering. In such a rescattering the leading hadron will produce new secondary particles and, as a consequence, its value of $x_L$ will be diminished. Nowadays this effect is called the rapidity gap survival factor $S^2$, since the secondaries populate and destroy the rapidity gap, $\Delta\eta \simeq$ ln $1/(1-x_L)$, between the leading hadron $c$ and the hadrons in the system $Y$. Since the $S^2$ was not accounted for explicitly, the determinations give only an {\it effective} value of the triple-Pomeron vertex, which implicitly embodies an $S^2$ factor. However the probability of soft rescattering, and hence the $S^2$ factor, depends on the nature of the incoming and leading hadrons. Moreover the value of $S^2$ depends on the kinematical variables of the process \cite{KKMRs}. Clearly it is important to find a way to measure the universal {\it bare} triple-Pomeron vertex, free from these rescattering effects.

A good illustration of the situation is an analogous problem which arises in the analysis of leading neutron production at HERA \cite{KKMRln,KMRln}. There, to understand the process we need to allow for rescattering (absorptive) corrections to pure Regge $\pi$ exchange. The rescattering modifies the predictions of not only the absolute value, but also the $t$ dependence of leading neutron production. Quantitatively it reduces the cross section by a factor of 0.4, and changes the $t$ slope, $B$ by 1 $\GeV^{-2}$, where $d\sigma /dt \sim {\rm exp}(Bt)$.

The $t$ dependence of the triple-Pomeron vertex plays a crucial role in the solution of the so-called Finkelstein-Kajantie problem \cite{fk,FK}. That is, if we neglect the $t$ dependence of the vertex and the gap survival factor, then it turns out, even in the case $\alpha_P(0)=1$, that the cross section for multigap events grows as a power of the energy, and so violates the Froissart limit. Each additional gap brings a ln$s$ factor arising from the integral over the gap size. The sum of these ln$s$ factors leads to the power behaviour.

Two solutions were proposed. First, the {\it weak coupling} solution in which the vertex {\it vanishes} as $t \to 0$ \cite{GM1}. In this case the ln$s$ factor arising from the gap size is compensated by a 1/ln$s$ caused by the shrinkage of the $t$ distribution with increasing energy. Interestingly, it predicts that, at ultra high energies, the total cross sections $\sigma_{ab}$ will tend to a universal constant independent of $a$ and $b$ \cite{G}. An alternative possibility is to suppress the multigap events by a strong absorptive correction, that is by decreasing the gap survival factors. In other words a large multigap cross section, calculated using a bare triple-Pomeron vertex which is {\it non-vanishing} as $t \to 0$, is multiplied by a gap survival factor which decreases faster with energy than the bare cross section. This leads to the Froissart-like black disc limit, and is called the {\it strong coupling} solution \cite{GM2}.

The old hadron-hadron data of (\ref{eq:hh}) show no indication that the triple-Pomeron vertex vanishes as $t \to 0$. However these data are inconclusive regarding the $t$ behaviour. The reasons are as follows. First, if it were to vanish, this behaviour would only be apparent at rather small $t$, $|t| \lapproxeq 1/R^2$, where $R$ is the typical transverse size of the incoming hadrons. Unfortunately, the data are not precise enough in this small $t$ region. Second, the effect is masked by the Pomeron cut contribution---soft rescattering washes out the momentum transferred through the individual Pomeron, and zero $p_t$ of the leading hadron does not mean that the momentum transfer in the bare triple-Pomeron vertex is $< 1/R^2$. It has been shown \cite{ab} that after accounting for multi-Pomeron effects, the old hadron-hadron data cannot eliminate the possibility that the bare triple-Pomeron vertex vanishes as $t \to 0$. 

\begin{figure}
\begin{center}
\includegraphics[height=5cm]{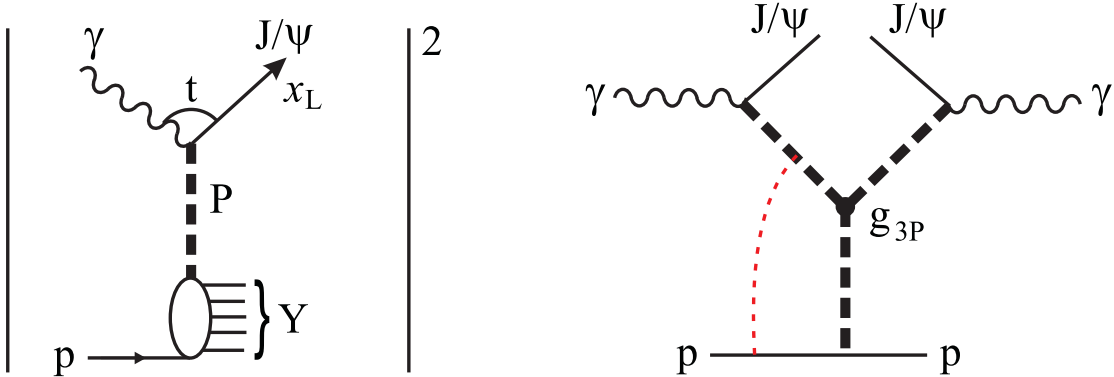}
\caption{The process of proton dissociation in diffractive $J/\psi$ photoproduction, $\gamma+p \to J/\psi+Y$, which is described by a diagram with a triple-Pomeron vertex in which the rescattering effects are small. The dotted line would mean the diagram became an enhanced diagram---this contribution is small. }
\label{fig:2}
\end{center}
\end{figure}

It is clearly important to study the triple-Pomeron interaction in a process where the rescattering effects are absent or strongly suppressed. Such a process, which is experimentally accessible, is proton dissociation in diffractive $J/\psi$ photo- ( or electro-) production
\be
\gamma ~+~p ~\to~ J/\psi ~+~Y.
\label{eq;psi}
\ee
The process is shown schematically in Fig.~\ref{fig:2}. By detecting the $\J$ in the final state, we select the charm component of the photon wave function, which has a small size of $\sim 1/m_c$. Such a component has a small absorptive cross section. Therefore the probability of an additional rescattering is much weaker. Even in the case of a nuclear target the probability of $\J$ rescattering is small \cite{resc,Kop}. Of course, there may be some `enhanced' absorptive corrections (of the type indicated by the dotted Pomeron-exchange line on the diagram on the right-hand-side of  Fig.~\ref{fig:2}). However the phenomenological analysis of leading neutron data \cite{KKMRln} demonstrates that such corrections are small at the available HERA energies. Otherwise the probability to observe a leading neutron would strongly depend on the initial $\gamma p$ energy, $W_{\gamma p}$, in contradiction with the data. Moreover, to provide enough phase space for such an `enhanced' correction, we would need to have sufficiently large rapidity intervals (either side of the dotted line) between the different vertices \cite{KMRenh}. This corresponds to an extremely small value of $1-x_L \lapproxeq 10^{-2}$, which is beyond the reach of the present experiments.

Another advantage of the process $\gamma+p \to J/\psi+Y$ is that, thanks to the small size of the $\J$ meson, the vanishing of the triple-Pomeron vertex (if true) would reveal itself over a more extended region of $t$. This possibility is practically already eliminated by the present data, see Fig. 4 of \cite{zeus}. Thus we can conclude that the strong coupling solution of the triple-Pomeron interaction is confirmed by the HERA $\J$ data.

Though the process $\gamma+p \to J/\psi+Y$ is experimentally accessible at HERA, so far no dedicated measurements are available in the literature. However some information does exist\footnote{We thank Michele Arneodo and Alessia Bruni for discussions concerning the data.}, as the process has been considered as a background to elastic $\J$ production, $\gamma+p \to J/\psi+p$. This somewhat fragmentary information allows us to make the following observations. We use $\sigma_Y$ and $\sigma_{\rm el}$ to denote the two $\J$ production cross sections.

\begin{itemize}
\item The energy dependence of $\J$ production with proton dissociation and for the elastic process are consistent with each other. If we write
\be
\sigma_i~\propto~W^{\delta_i}
\ee
where $W$ is the photon-proton centre-of-mass energy, then \cite{zeus695,h110,zeus24,Akt}
\begin{eqnarray}
\delta_Y~£\simeq£~0.7\pm 0.2  \nonumber \\
\delta_{\rm el}~£\simeq£~0.7 \pm 0.05.
\label{eq:el}
\end{eqnarray}
Next, the $M_Y$ dependence \cite{zeus695} for a fixed value of the ratio $M_Y^2/W^2$ is consistent with the behaviour of hadronic total cross sections, that is
\be
\sigma_Y \propto (M_Y^2)^\epsilon~~~~~~{\rm with} ~~\epsilon \simeq 0.08.
\label{eq:tot}
\ee
The above results are the properties expected from the triple-Pomeron term\footnote{Note that the Pomeron trajectories $\alpha_P(0)$ and $\alpha_P(t)$ in (\ref{eq:trf}), that is in the triple-Pomeron diagram in Fig.~\ref{fig:2}, are not the same. The lower Pomeron $\alpha_P(0)$ in Fig.~\ref{fig:2} is the usual `soft' Pomeron; whereas the upper ones, with $\alpha_P(t)$,
include DGLAP evolution from a low initial scale $\mu=\mu_0$ up to a rather large scale $\mu\sim M_{J/\psi}$ at the $J/\psi$
production vertex. The summation of the double logarithms $(\alpha_s\ln(1/x)\ln(\mu^2/\mu^2_0))^n$ leads to a steeper $x$-dependence and hence to a larger effective intercept
for the trajectory $\alpha_P(t)$ of the upper `hard' Pomeron. Thus in  (\ref{eq:el}) and (\ref{eq:tot}) we have $\delta/4 >
\epsilon$, where $\delta=4(\alpha^{\rm hard}_P(0)-1)$ and
$\epsilon=\alpha^{\rm soft}_P(0)-1$.} in (\ref{eq:trf}).
\item
The $t$ slope of $\J$ production by proton dissociation is observed \cite{zeus24,zeus695} to be rather small, $b_Y \simeq 0.7 \pm 0.2 ~\GeV^{-2}$. In some sense, we can regard $b_Y$ to be driven by the Pomeron form factor. On the other hand, the slope for elastic $\J$ production, which is driven by the proton form factor, is measured \cite{h1472,h175} to be $b_{\rm el} \simeq 4.5~ \GeV^{-2}$. This indicates that the size of the Pomeron and the triple-Pomeron vertex are much less than that for the proton. This conclusion about the small size of the triple-Pomeron vertex is confirmed by the data on $\rho$ production, where the difference of slopes, $\Delta b = b_{\rm el}-b_Y\simeq 4~\GeV^{-2}$ \cite{zeus818}. These properties justify the idea that the pure triple-Pomeron coupling may be extracted from a triple-Regge analysis of data for $\J$ production with proton dissociation. The small size of the Pomeron indicates that the rescattering corrections are indeed small. This information comes from a consideration of proton dissociation as a background process in the presentation of elastic $\J$ production data. However there is more detailed information on the $t$ dependence of the proton dissociation process in Fig. 4 of \cite{zeus}, that we mentioned previously.
\item
A comparison of the cross sections is also revealing. For $\J$ production, an approximate estimate gives \cite{zeus,zeus24}
\be
\sigma_Y / \sigma_{\rm el} ~\simeq~1 \pm 0.3,
\ee
while for $\rho$ production we have \cite{h175,zeus2}
\be
\sigma_Y / \sigma_{\rm el} ~\simeq~0.6 \pm 0.2.
\ee
This difference in the values of the ratios is anticipated because strong absorption is expected in the case of $\sigma_Y$ for $\rho$ production.
\end{itemize}

The above summary of the existing, admittedly fragmentary, data on $\gamma+p \to J/\psi+Y$ certainly supports a dominant triple-Pomeron behaviour. So we may attempt an extraction of the {\it bare} triple-Pomeron coupling, $g_{3P}(0)$. We assume that the data for $\sigma_Y$, collected \cite{zeus} in the interval\footnote{We thank Alessia Bruni for informing us about the interval of $M_Y$.} of $M_Y$ from 2.5 GeV up to $M_Y^2=0.1 W^2$, is described just by the triple-Pomeron term in (\ref{eq:trf}). Then
\be
\int dM_Y^2 \frac{d\sigma_Y}{dt dM_Y^2}~~/~~\frac{d \sigma_{\rm el}}{dt}~~=~~\frac{1}{16\pi^2}g^{\rm bare}_{3P}(0)\cdot J~~/~~\frac{1}{16\pi}g_N(0),
\ee
where we estimate that the integration over $M_Y$ gives the factor $J \simeq 0.6$. In this way we find
\be
g^{\rm bare}_{3P}(0)~/~g_N(0) ~~\simeq~~ 1~/~3.
\label{eq:bare}
\ee
This has to be compared with the ratio extracted \cite{old} many years ago from hadron-hadron initiated data
\be
g^{\rm effective}_{3P}(0)~/~g_N(0) ~~\simeq~~ 1~/~10.
\ee
The distortion of the coupling in the old estimate, due to the survival factors, is clearly evident\footnote{Interestingly, a similar distortion had been already allowed for in the global analysis of `soft' hadronic data that was performed in \cite{KMRsoft}.}.
   The fact that including absorptive effects gives a larger value of the triple-Pomeron coupling, is not new. Larger values of $g_{3P}$ were obtained, for example, in \cite{ab,large}.
   However all the previous estimates are model dependent. We cannot sum up all the Reggeon diagrams which describe the absorptive effects. Therefore the value of $g_{3P}$ depends on the class of multi-Pomeron diagrams chosen for resummation, on the assumptions about the behaviour of the multi-Pomeron vertices, and sometimes on the threshold factor in the multi-Pomeron vertices, etc.  The advantage of the present evaluation of $g_{3P}$ is that we study a process where the absorptive corrections are small, and so affect the final result much less.

Of course, in the integration over $M_Y$, we cannot neglect the possible contributions of secondary Reggeons in (\ref{eq:trf}), such as the RRP term near the upper limit of the integral and the PPR term for $M_Y$ close to $M_Y$(min). The value of $g^{\rm bare}_{3P}(0)$ is therefore expected to be a bit smaller than in (\ref{eq:bare}). It is thus crucially important to have $\gamma+p \to J/\psi+Y$ data with an explicit measurement of the $M_Y$ spectrum in order to perform a full triple-Regge analysis in which we quantify the different triple-Regge contributions.

The discussion in this paper has implications for all diffractive processes. We mention two topical examples, both of which illustrate the importance of the energy dependence of the survival factor $S^2$.  In the existing Monte Carlos for high energy multiparticle production, the distributions of leading particles are parametrized in the triple-Regge form. In comparison with the limiting fragmentation or Feymann scaling hypothesis \cite{limitingfrag,collins}, where the normalized cross section does not depend on energy, we now have
\be
\frac{1}{\sigma_{\rm tot}} \frac{d\sigma^{\rm SD}}{dt dx_L}~~=~~\frac{g_N^2(t) g_{3P}(t)}{16\pi^2 g_N(0)} ~(1-x_L)^{\alpha_P(0)-2\alpha_P(t)}~S_{\rm SD}^2(s,t),
\label{eq:sss}
\ee
which depends, not only on $x_L$, but on the energy\footnote{Note, however, that as it stands,
formula (\ref{eq:sss}) does not allow for the effects of migration which, as we have
seen in the case of the leading neutrons \cite{KKMRln}, give a noticeable contribution for
  $x_L<0.7-0.8$. Therefore (\ref{eq:sss}) is only valid in its present form for $x_L \gapproxeq 0.8$. Also (\ref{eq:sss}) contains just the PPP-contribution. For $x_L<0.8-0.9$ allowance should be made for a RRP contribution
with its own $S^2_{\rm RRP}$.}. The energy dependence comes entirely from the survival factor $S_{\rm SD}^2$. So, before we discuss the two examples, we compute the energy dependence of $S^2_{\rm SD}$. As before, we use SD to denote single-particle diffractive dissociation.

\begin{figure}[t]
\begin{center}
\includegraphics[height=12cm]{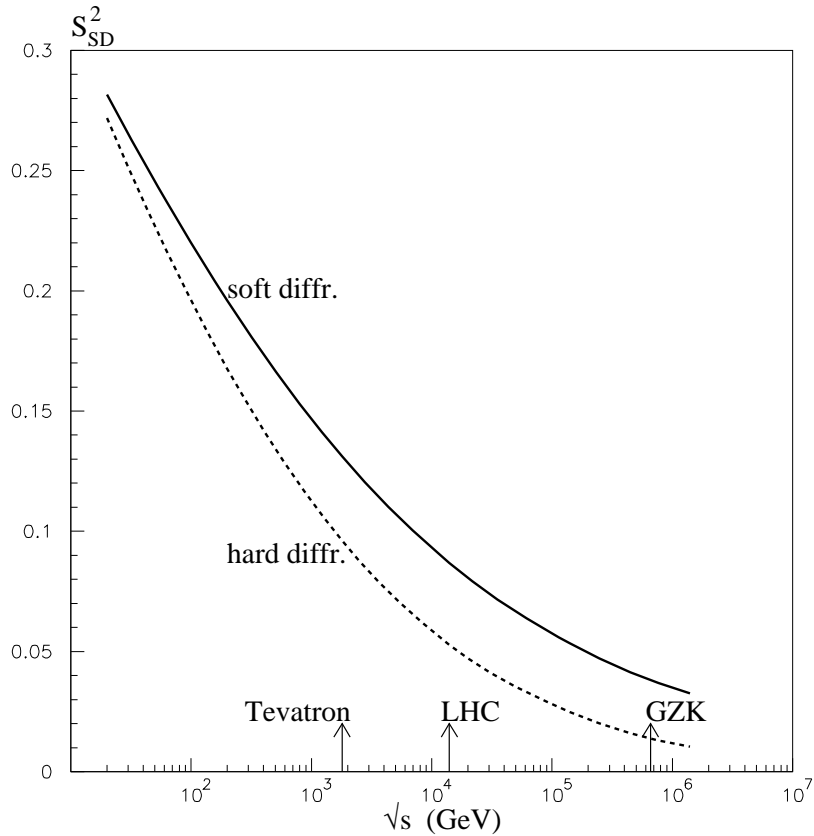}
\caption{The energy dependence of the survival factor, $S^2_{SD}$, for single (soft and hard) diffractive dissociation, via the triple-Pomeron interaction.  The GZK energy, $2.3 \times 10^{20}$ eV, is the theoretical upper (Greisen-Zatsepin-Kuzmin \cite{GZK}) limit on the energy of cosmic rays which arises from the large cross section for the interaction of the protons with relic photons due to $\Delta$ isobar production.}
\label{fig:3}
\end{center}
\end{figure}
In Fig.~\ref{fig:3} we show the expected energy dependence of the survival factor $S_{\rm SD}^2$ for the cross section integrated over $t$. This was calculated in a simplified two-channel eikonal model\footnote{The details of the two-channel eikonal model will be presented elsewhere.} with the $t$ dependence of the elastic Pomeron-proton vertex, $g_N(t)$, given by the proton electromagnetic form factor $F_1$, that is $g_N(t)=g_N(0)F_1(t)$. The parameters of the Pomeron trajectory were tuned to describe the CERN-ISR, CERN-SPS and Tevatron elastic data. It is clear from the figure that the energy dependence of $S_{\rm SD}^2$ is quite pronounced.  Moreover, we find that the predictions for $S_{\rm SD}^2$ obtained using a two-channel eikonal are stable\footnote{If we try to mimic the two-channel eikonal by an `enhanced' one-channel eikonal, then the predictions for $S_{\rm SD}^2$ decrease and the values at ultra high energies become too low and unstable.} to changing the details of the model, after the `soft' parametrization is tuned to fit the elastic data.

The continuous curve in Fig.~\ref{fig:3} is for {\it soft} diffractive dissociation. Here the process takes place at relatively large impact parameters $b_t$, where the absorption is not so strong. Suppose, instead, we were to consider {\it hard} diffractive dissociation, for example where a pair of high $E_T$ jets \cite{KKMRs} or a heavy quark pair or $W$ or $Z$ boson or some other heavy object is produced within the system $Y$. In these cases the interaction takes place at much smaller $b_t$. Here, in the central $b_t$ region, the absorption is stronger and the corresponding $S_{\rm SD}^2$ factor is smaller, especially at ultra high energies where the proton opacity becomes very close to the black disc limit. Therefore, for completeness, we also show the predictions for the energy dependence of $S_{\rm SD}^2$ for hard diffractive dissociation, by the dashed curve in Fig.~\ref{fig:3}.

Our first topical example, where the energy dependence of $S_{\rm SD}^2$ plays an important role, is the evaluation of the so-called pile-up backgrounds to forward physics studies at the LHC, see, for instance, \cite{alb}. At high LHC luminosities, the soft pile-up single (and even double) diffractive processes can fake the signal from hard diffractive events due to overlap with the non-diffractive hard processes observed in the central detector. However, when we allow for the energy dependence of $S_{\rm SD}^2$ (for soft diffraction), we decrease the predicted leading proton spectra at the LHC energies, as compared to the leading proton spectrum measured at HERA \cite{zeuslps}.
A preliminary estimate\footnote{The reduction by about 3 is only an illustrative
factor, as it is based on
$S^2 \sim 0.4$ found \cite{KMRln} for leading neutrons at HERA. For leading protons the value of $S^2$
may be a bit smaller, since a Pomeron-exchange interaction has smaller $b_t$
than for $\pi$-exchange leading neutron
production. To obtain a quantitative estimate of the reduction factor it will be necessary to perform a careful analysis
of the leading protons produced in DDIS \cite{zeuslps}. } gives a reduction by about a factor of 3.
Therefore the impact of pile-up is be suppressed by the energy dependence of $S_{\rm SD}^2$. 

The second example illustrates the importance of the energy dependence of $S_{\rm SD}^2$ to the analysis of ultra high energy cosmic ray data. The observed structure of the extensive cosmic ray air showers depends sensitively on the spectrum of the leading hadrons. From the continuous curve in Fig.~\ref{fig:3} we see that in going from HERA energies to the GZK cut-off energy \cite{GZK}, the survival probability $S_{\rm SD}^2$ decreases by almost an order of magnitude.  Of course, the small values of $S_{\rm SD}^2$ at ultra high energies do not mean that the leading hadron disappears. Instead, after the rescattering, the large $x_L$ hadron migrates to a lower $x_L$ and larger $p_t$ domain \cite{KKMRln}, thereby changing the shape of the extensive air shower.

\section*{Acknowledgements}

We thank Michele Arneodo, Alessia Bruni and Risto Orava for valuable discussions. MGR and VAK thank, respectively, the IPPP at the University of Durham and the Aspen Center for Physics for hospitality. This work was supported by the Royal Society,
by INTAS grant 05-103-7515, by grant RFBR 04-02-16073 and by the Federal Program of the Russian Ministry of Industry, Science and Technology SS-1124.2003.2.


\end{document}